\documentclass[10pt,twocolumn]{IEEEtran}
\usepackage{amssymb}
\usepackage{amsfonts}
\usepackage{amsmath}
\usepackage{algorithm}
\usepackage{graphicx,subfigure,cite,multicol,multirow,diagbox,booktabs,array}

\usepackage[dvips]{color}
\usepackage{float}
\usepackage[noend]{algpseudocode}
\ifCLASSINFOpdf
\else
\fi
\begin{document}
\title{Two Efficient Beamformers for Secure Precise Jamming and Communication with Phase Alignment
}
\author{Feng Shu,~\emph{Member},~\emph{IEEE},~Lingling Zhu,~Wenlong Cai,\\Tong~Shen,~Jinyong Lin,~Shuo Zhang,~and Jiangzhou Wang,~\emph{Fellow},~\emph{IEEE}
\thanks{This work was supported in part by the National Natural Science Foundation of China (Nos. 61771244, 61501238, 61702258, 61472190, and 61271230)(Corresponding authors: Feng Shu, Wenlong Cai, Jinyong Lin).}
\thanks{Feng Shu,~Lingling Zhu,~and~Tong Shen,~and are with the School of Electronic and Optical Engineering, Nanjing University of Science and Technology, 210094, CHINA. (Email: shufeng@njust.edu.cn).}
\thanks{Wenlong Cai,~Jinyong Lin,~and Shuo Zhang are with Beijing Aerospace Automatic Control Institute, Beijing 100854, China. (Email: caiwenlon@buaa.edu.cn).}
\thanks{Jiangzhou Wang is with the School of Engineering and Digital Arts, University of Kent, Canterbury CT2 7NT, U.K. Email: \{j.z.wang\}@kent.ac.uk.
}

}
\maketitle

\begin{abstract}
 To achieve a better effect of interference on eavesdropper with an enhanced security, a secure precise jamming (PJ) and communication (SPJC)  is proposed and its basic idea is to force the transmit energy of artificial noise (AN) and confidential message into the neighborhoods of Eve and Bob  by using random subcarrier selection (RSS), directional modulation, and beamforming under phase alignment (PA) constraint (PAC). Here, we propose two high-performance beamforming schemes: minimum transmit power (Min-TP) and  minimum regularized transmit power (Min-RTP) to achieve SPJC under PAC and orthogonal constraint (OC), where OC means that AN and CM are projected onto the null-spaces of the desired and eavesdropping channels, respectively. Simulation results show that the proposed Min-TP and Min-RTP methods perform much better than existing equal amplitude (EA) method in terms of both bit-error-rate (BER) and secrecy rate (SR) at medium and high signal-to-noise ratio regions. The SR performance difference  between the proposed two methods becomes  trivial as the number of transmit antennas approaches large-scale. More importantly, we also find the fact that all three schemes including EA, Min-TP, and Min-RTP can form two main peaks of AN and CM around Eve and Bob, respectively. This achieves both PJ and secure precise wireless transmission (SPWT), called SPJC.
\end{abstract}


\IEEEpeerreviewmaketitle

\section{Introduction}
\IEEEPARstart{R}{ecently}, physical-layer security (PLS) has become a promising research field in wireless communications and networking\cite{wyner,wanghuiming,zhaonan,chenxiaoming,wuyongpeng}. Several kinds of secure tools are developed for PLS such as relay cooperation, full-duplex, and secure modulations. To achieve a secure transmission in fading channel, secure spatial modulation is a proper choice by transmit antenna selection, beamforming of confidential message (CM), and artificial noise (AN) projection\cite{wangzhengwang,xiaguiyang,yuxiangbin}. However, for line-of-propagation (LoP) channel, directional modulation (DM)\cite{yuanding} has attracted recently substantial research activities. Specially, DM can be directly applied to millimeter wave and UAV channels. This makes it have a potential to become a hot research topic in the coming future and will provide an alternative secure solution for the future wireless networks, particularly beyond 5G and UAV networks\cite{Zhou2019UAV}.

As an enhanced secure version of DM, secure precise wireless transmission (SPWT) based on DM was proposed in \cite{huAN,wuxiaomin1}. In such a scheme, the range-angle-dependent characteristic is established and there is a single main peak of CM is formed around the desired receiver. The CM energy leakage outside the main peak can be omitted and seriously degraded by AN. In order to ensure the maximum secrecy rate (SR) of proximal Bob and eavesdropper, the authors in\cite{linjingran} optimized the frequency offsets of frequency diverse array by block successive upper-bound minimization algorithm. However, if Eve is sensitive enough, CM is easy to be intercepted by Eve. Therefore, the authors in \cite{qiubin} added AN to the proximal Bob and Eve scenario, and extended the frequency offset optimization in the case of multiple eavesdroppers. To reduce the complexity of SPWT receiver, a SPWT structure using random subcarrier selection (RSS) of orthogonal frequency diversion multiplexing (OFDM) is proposed \cite{wuxiaomin1}.  In \cite{shentong:journals/corr/abs-1808-01896}, the authors proposed two practical RSS methods to achieve a SWPT per OFDM symbol by randomization procedure including integer mod, ordering, and block interleaving.

 Since  a SWPT has been implemented, can we achieve a precise jamming (PJ) on Eve? In other words, a single AN energy main peak is formed at the position of Eve.  To address this issue,   the framework of secure PJ and communication (SPJC) is proposed and the corresponding two schemes  are developed. Our main contributions in this paper are as follows:
\begin{enumerate}
 \item  To impose a strong interference on Eve, we extend the idea of SPWT to form a framework of SPJC. By doing so, the transmit powers of AN and CM are focused at a small neighborhoods of Eve and Bob. Subsequently, a low-complexity beamforming method of minimizing transmit power (Min-TP) is proposed  subject to orthogonal constraint (OC) and  phase alignment constraint (PAC). The proposed Min-TP has a closed form expression. Simulation results show the proposed Min-TP performs much better than existing equal amplitude (EA) method.   However, the Gram matrix in the closed-form formula of Min-TP is singular. Instead, the pseudo-inverse of the Gram matrix  is used. This will lead to some performance loss.
 \item  To address the existing singular problem in the proposed Min-TP, a regularized method, called minimizing regularized transmit power (Min-RTP),  is proposed to provide an  improved robust solution, which is still under OC and PAC. Here, the optimal values of two regularized factors are obtained by two-dimensional (2D) exhaustive search. Simulation results show that, compared to EA and Min-TP, the proposed  Min-RTP perform better in the medium and high signal-to-noise ratio (SNR) regions in terms of SR and bit-error-rate (BER) performance.
     \end{enumerate}

The remainder is organized as follows. In Section II, we describe SPJC system model. Subsequently, we propose two beamforming methods: Min-TP and Min-RTP in Section III. In Section IV, the performance of the proposed scheme is numerically evaluated, and conclusions are given in Section V.

Notations: matrices, vectors, and scalars are denoted by letters of bold upper case, bold lower case, and lower case, respectively. Signs $(\cdot)^T$, $(\cdot)^H$, $(\cdot)^{\ast}$, $(\cdot)^{-1}$ and $(\cdot)^{\dag}$ denote matrix transpose, conjugate transpose, conjugate, inverse and Moore-Penrose pseudo inverse, respectively. The symbol $\mathbf{I}_N$ denotes the $N \times N$ identity matrix.
\section{System Model}
\begin{figure}[h]
  \centering
 \includegraphics[width=0.40\textwidth]{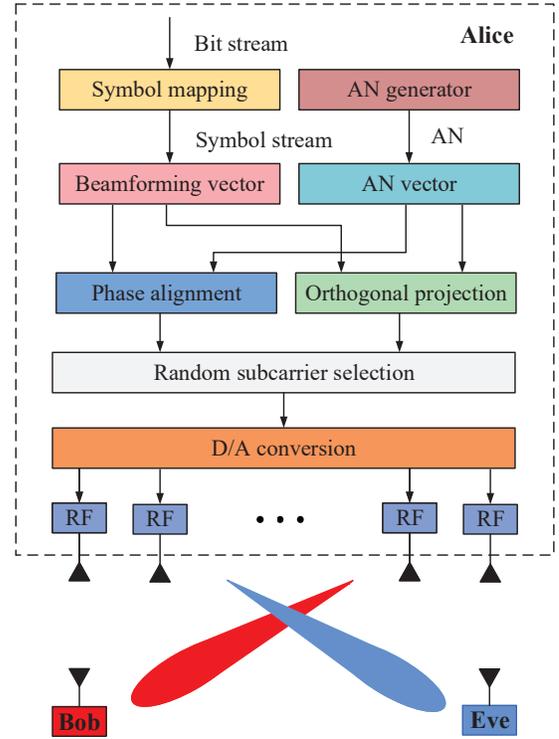}\\
  \caption{Block diagram for SPJC systems.}\label{Sys_Mod}
\end{figure}

Fig.~\ref{Sys_Mod} sketches a typical architecture for SPJC system. Here, an $N$-antenna uniform linear transmit array are employed at Alice, a single-antenna  at Bob and a single-antenna at Eve. AN and CM are transmitted to Eve and Bob via the same sequence of randomly-selected multiple subcarriers from all-subcarrier set of OFDM. The all-subcarrier set of OFDM is denoted as $S_{sub} = \{f_m|f_m=f_c+ m\Delta f,  m=0,1,\ldots,N_S-1\}$, where $f_c$ is the carrier frequency and $\Delta f$ is the subchannel bandwidth. The subcarrier assigned to $n$-th antenna is $f_n$, where $f_n \in S_{sub}$\cite{wuxiaomin1}.

In what follows, it is assumed that the channel model is LoP. The normalized steering vector for the transmit antenna array is as follows
\begin{align}
\mathbf{h}(\theta,R) = \frac{1}{\sqrt{N}}[e^{j \Psi_0(\theta,R)},\cdots, e^{j \Psi_{N-1}(\theta,R)}]^T,
\end{align}
where $\Psi_n(\theta,R) = 2\pi(f_c+k_n\Delta f)\frac{R-(n-1)d\cos\theta}{c} - 2\pi f_c\frac{R}{c}$ with $d$ being element spacing of uniform linear array (ULA) and $c$ being light speed. In general, frequency increment and central carrier frequency are required to satisfy $N_S\Delta f  \ll f_c$. In the following, with a high-resolution direction of arrival estimation\cite{Qinyaolu}, the corresponding steering vectors for Bob and Eve are $\mathbf{h}(\theta_B,R_B)$ and $\mathbf{h}(\theta_E,R_E)$, respectively, where $\theta_B$ and $\theta_E$ are the directional angles of Bob and Eve while $R_B$ and $R_E$ are the distances from Alice to Bob and Eve.

The baseband transmit signal is given by
\begin{align}
\mathbf{s} = \mathbf{v}_{CM}x + \mathbf{v}_{AN}z,
\end{align}
where $x$ is the CM  and $z$ is the AN both with average power constraint (i.e., $E[\left | x \right |^2]=1$, $E[\left | z \right |^2]=1$). $\mathbf{v}_{CM} $ and $\mathbf{v}_{AN}$ are the CM and  AN beamforming vectors, respectively.

Accordingly, the received signals at position of Bob and Eve are
\begin{align}
y(\theta_B,R_B)
&=\sqrt{g_b}\mathbf{h}^H(\theta_B,R_B)\mathbf{v}_{CM}x \\
&+\sqrt{g_b}\mathbf{h}^H(\theta_B,R_B)\mathbf{v}_{AN}z+n_B\nonumber,
\end{align}
and
\begin{align}
y(\theta_E,R_E)
&=\sqrt{g_e}\mathbf{h}^H(\theta_E,R_E)\mathbf{v}_{CM}x \\
&+\sqrt{g_e}\mathbf{h}^H(\theta_E,R_E)\mathbf{v}_{AN}z+n_E\nonumber,
\end{align}
respectively, where $g_b = g_0R_B^{-2}$ and $g_e = g_0R_E^{-2}$ denote the path loss coefficients from Alice to Bob and from Alice to Eve. $g_0$ is the reference distance and set to 1m. $n_B$ and $n_E$ are the additive white Gaussian noise (AWGN) with distribution $n_B\sim\mathcal{C}\mathcal{N}(0,\sigma_B^2)$ and $n_E\sim\mathcal{C}\mathcal{N}(0,\sigma_E^2)$, respectively.
\section{Two Proposed Beamforming Schemes with OC and PAC}
In this section, under the OC and PAC, the two novel beamforming schemes Min-TP and Min-RTP are proposed to achieve the SPJC.  Also, we derive their approximate closed-form expressions by the Lagarangian multiplier. The penalty factors or regression coefficients of the latter are attained by 2D exhaustive search.
\subsection{Proposed Min-TP}
Now, we turn to the optimization design of Max-TP. The corresponding optimization problem of Max-TP is casted as
\begin{align}\label{P1}
&\min_{\mathbf{v}_{CM}}~~~~~~~~~~~~~~~~~\mathbf{v}_{CM}^H\mathbf{v}_{CM}\\
&~~\text{s.t.} ~~~~(\text{OC})~~~~~~~\mathbf{h}^H(\theta_E,R_E)\mathbf{v}_{CM}=0 \nonumber\\
&~~~~~~~~~(\text{PAC})~~~~~~\mathbf{h}^H(\theta_B,R_B)\mathbf{v}_{CM}=1,\nonumber
\end{align}
where the first constraint (OC) $\mathbf{h}^H(\theta_E,R_E)\mathbf{v}_{CM}=0$ is to force CM to transmit along the null-space (NS) of Eve and similarly the second constraint (PAC) $\mathbf{h}^H(\theta_B,R_B)\mathbf{v}_{CM}=1$ forces the CM signal transmit along the desired direction. To simplify the above optimization problem, the first constraint OC implies that the  $\mathbf{v}_{CM}$ is represented as
\begin{align}
\mathbf{v}_{CM}=(\mathbf{I}_N - \mathbf{h}(\theta_E,R_E)\mathbf{h}^H(\theta_E,R_E))\mathbf{u}_{CM}£¬
\end{align}
with $\mathbf{u}_{CM}$ as a new optimization variable. Substituting the above equation in (\ref{P1}) yields the  following simplified optimization problem as
\begin{align}\label{P2}
&\min_{\mathbf{u}_{CM}}~~~~~~~~~~~~~\mathbf{u}_{CM}^H(\mathbf{I}_N - \mathbf{h}(\theta_E,R_E)\mathbf{h}^H(\theta_E,R_E))^H\nonumber\\
&~~~~~~~~~~~~~~~~\cdot(\mathbf{I}_N - \mathbf{h}(\theta_E,R_E)\mathbf{h}^H(\theta_E,R_E))\mathbf{u}_{CM}\\
&~\text{s.t.} ~~~(\text{PAC})~~~\mathbf{h}^H(\theta_B,R_B)(\mathbf{I}_N- \mathbf{h}(\theta_E,R_E)\mathbf{h}^H(\theta_E,R_E)) \nonumber\\
&~~~~~~~~~~~~~~~~~\cdot\mathbf{u}_{CM}=1,\nonumber
\end{align}
where the OC has been removed. Let us define $\mathbf{A} = \mathbf{I}_N - \mathbf{h}(\theta_E,R_E)\mathbf{h}^H(\theta_E,R_E)$, and $\mathbf{v}_{CM} = \mathbf{A}\mathbf{u}_{CM}$, then the optimization problem in  (\ref{P2}) can be rewritten as
\begin{align}\label{P3}
&\min_{\mathbf{u}_{CM}}~~~~~~~~~~~~~\mathbf{u}_{CM}^H\mathbf{A}^H\mathbf{A}\mathbf{u}_{CM}\\
&~\text{s.t.}~~~~(\text{PAC})~~~\mathbf{h}^H(\theta_B,R_B)\mathbf{A}\mathbf{u}_{CM}=1.\nonumber
\end{align}
The Lagrangian function associated with the above optimization is defined as
\begin{align}\label{P6}
L(\mathbf{u}_{CM}, \lambda) = \mathbf{u}_{CM}^H\mathbf{A}^H\mathbf{A}\mathbf{u}_{CM} + \lambda (\mathbf{h}^H(\theta_B,R_B)\mathbf{A}\mathbf{u}_{CM} - 1).
\end{align}
If ($\mathbf{u}_{CM}^\star, \lambda$) is the optimal solution to the above equation, the first-order derivative of the Lagrangian function should be set to be zero as follows
\begin{align}
\frac{\partial L(\mathbf{u}_{CM}, \lambda)}{\partial(\mathbf{u}_{CM})} = (\mathbf{A}^H\mathbf{A})^{T}\mathbf{u}_{CM}^{*} + \lambda(\mathbf{h}^H(\theta_B, R_B)\mathbf{A})^{T}=0,
\end{align}
which gives directly
\begin{align}\label{P4}
\mathbf{u}_{CM} = -\lambda(\mathbf{A}^H\mathbf{A})^{\dag}\mathbf{A}^{H}\mathbf{h}(\theta_B, R_B).
\end{align}
Inserting (\ref{P4}) back into the PAC $\mathbf{h}^H(\theta_B,R_B)\mathbf{A}\mathbf{u}_{CM}=1$, the Lagrange multiplier can be calculated as
\begin{align}
\lambda=\frac{-1}{\mathbf{h}^H(\theta_B,R_B)\mathbf{A}(\mathbf{A}^H\mathbf{A})^{\dag}\mathbf{A}^H\mathbf{h}(\theta_B,R_B)}.
\end{align}
Substituting the above back into (\ref{P4}), we have
\begin{align}
\mathbf{u}_{CM}^\star =\frac{(\mathbf{A}^H\mathbf{A})^{\dag}\mathbf{A}^{H}\mathbf{h}_B(\theta_B, R_B)}{\mathbf{h}^H(\theta_B,R_B)\mathbf{A}(\mathbf{A}^H\mathbf{A})^{\dag}\mathbf{A}^H\mathbf{h}(\theta_B,R_B)}.
\end{align}
Therefore, the optimal value of $\mathbf{v}_{CM}$ is given by
\begin{align}
\mathbf{v}_{CM} =\frac{\mathbf{A}(\mathbf{A}^H\mathbf{A})^{\dag}\mathbf{A}^{H}\mathbf{h}(\theta_B, R_B)}{\mathbf{h}^H(\theta_B,R_B)\mathbf{A}(\mathbf{A}^H\mathbf{A})^{\dag}\mathbf{A}^H\mathbf{h}(\theta_B,R_B)}.
\end{align}
In the same manner, the AN beamforming vector  $\mathbf{v}_{AN}$ is optimized by minimizing the transmit power of AN subject to two corresponding constraints OC and PAC, i.e.,
\begin{align}\label{P5}
&\min_{\mathbf{v}_{AN}}~~~~~~~~~~~~~~~\mathbf{v}_{AN}^H\mathbf{v}_{AN}\\
&~\text{s.t.} ~~~~~(\text{OC})~~~~  \mathbf{h}^H(\theta_B,R_B)\mathbf{v}_{AN}=0\nonumber\\
&~~~~~~~~~(\text{PAC})~~~\mathbf{h}^H(\theta_E,R_E)\mathbf{v}_{AN}=1,\nonumber
\end{align}
Similar to the optimization process of $\mathbf{v}_{CM}$, $\mathbf{v}_{AN}$ is given by
\begin{align}
\mathbf{v}_{AN} =\frac{\mathbf{B}(\mathbf{B}^H\mathbf{B})^{\dag}\mathbf{B}^{H}\mathbf{h}(\theta_E, R_E)}{\mathbf{h}^H(\theta_E,R_E)\mathbf{B}(\mathbf{B}^H\mathbf{A})^{\dag}\mathbf{B}^H\mathbf{h}(\theta_E,R_E)},
\end{align}
where $\mathbf{B} = \mathbf{I}_{N} - \mathbf{h}(\theta_B,R_B)\mathbf{h}^H(\theta_B,R_B)$ can project AN onto the NS of $\mathbf{h}^H(\theta_B, R_B)$.

\subsection{Proposed Min-RTP}
Since the Gram matrix $\mathbf{A}$ and $\mathbf{B}$ in Subsection A are singular matrices, the solutions to $\mathbf{v}_{CM}$ and $\mathbf{v}_{AN}$  are obtained by using the pseudo-inverse operation. This will result in some stable problem or performance loss. To address this problem,   a regularized penalty is added to the objective function, the optimization problem in (\ref{P3}) can be converted into
\begin{align}\label{P7}
&\min_{\mathbf{u}_{CM},\gamma}~~~~~~~~~~\mathbf{u}_{CM}^H\mathbf{A}^H\mathbf{A}\mathbf{u}_{CM} + \gamma_{CM}\mathbf{u}_{CM}^H\mathbf{u}_{CM}\\
&~\text{s.t.}~~~~(\text{PAC})~~\mathbf{h}^H(\theta_B,R_B)\mathbf{A}\mathbf{u}_{CM}=1,\nonumber
\end{align}
where $\gamma_{CM}\mathbf{u}_{CM}^H\mathbf{u}_{CM}$ is the regularized term and $\gamma_{CM}$ is the associated regularization factor. Similar to (\ref{P6}), we have
\begin{align}
L(\mathbf{u},\gamma_{CM}) &= \mathbf{u}_{CM}^H\mathbf{A}^H\mathbf{A}\mathbf{u}_{CM} + \gamma_{CM}\mathbf{u}_{CM}^H\mathbf{u}_{CM}\nonumber\\
&+ \lambda_R (\mathbf{h}^H(\theta_B,R_B)\mathbf{A}\mathbf{u}_{CM} - 1),
\end{align}
where $\lambda_R$ is the Lagrange multiplier. The optimum solution $\mathbf{u}_{CM}^\star$ can be obtained by setting  the partial derivative of the Lagrange function $L$ equal 0, i.e.,
\begin{align}
\frac{\partial L(\mathbf{u}_{CM},\gamma_{CM})}{\partial(\mathbf{u}_{CM})} &= (\mathbf{A}^H\mathbf{A} + \gamma_{CM}\mathbf{I}_N)^{T}\mathbf{u}_{CM}^{*}\nonumber\\
&+ \lambda_R(\mathbf{h}^H(\theta_B, R_B)\mathbf{A})^{T}=0,
\end{align}
which yields
\begin{align}
\mathbf{u}_{CM} = -\lambda_R((\mathbf{A}^H\mathbf{A}+\gamma_{CM} \mathbf{I}_N )^{-1})^{\ast}\mathbf{A}^{H}\mathbf{h}(\theta_B, R_B),
\end{align}
and
\begin{align}
\lambda_R=\frac{-1}{\mathbf{h}^H(\theta_B,R_B)\mathbf{A}((\mathbf{A}^H\mathbf{A}+\gamma_{CM}\mathbf{I}_N)^{-1})^{\ast}\mathbf{A}^H\mathbf{h}(\theta_B,R_B)},
\end{align}
respectively, which gives the following $\mathbf{u}_{CM}$ and $\mathbf{v}_{CM}$
\begin{align}
\mathbf{u}_{CM} =\frac{((\mathbf{A}^H\mathbf{A}+\gamma_{CM} \mathbf{I}_N )^{-1})^{\ast}\mathbf{A}^{H}\mathbf{h}(\theta_B, R_B)}{\mathbf{h}^H(\theta_B,R_B)\mathbf{A}((\mathbf{A}^H\mathbf{A}+\gamma_{CM}\mathbf{I}_N)^{-1})^{\ast}\mathbf{A}^H\mathbf{h}(\theta_B,R_B)},
\end{align}
and
\begin{align}
\mathbf{v}_{CM}= \frac{\mathbf{A}((\mathbf{A}^H\mathbf{A}+\gamma_{CM} \mathbf{I}_N )^{-1})^{\ast}\mathbf{A}^{H}\mathbf{h}(\theta_B, R_B)}{\mathbf{h}^H(\theta_B,R_B)\mathbf{A}((\mathbf{A}^H\mathbf{A}+\gamma_{CM}\mathbf{I}_N)^{-1})^{\ast}\mathbf{A}^H\mathbf{h}(\theta_B,R_B)}.
\end{align}

Likewise, $\mathbf{u}_{AN}$ is computed  by optimizing
\begin{align}
&\min_{\mathbf{u}_{AN}}~~~~~~~~~~~~~~\mathbf{u}_{AN}^H\mathbf{B}^H\mathbf{B}\mathbf{u}_{AN} + \gamma_{AN}\mathbf{u}_{AN}^H\mathbf{u}_{AN}\\
&~\text{s.t.} ~~~~~(\text{PAC})~~~ \mathbf{h}^H(\theta_E,R_E)\mathbf{B}\mathbf{u}_{AN}=1.\nonumber
\end{align}
In a similar way, we have the optimal $\mathbf{v}_{AN}$  as
\begin{align}
\mathbf{v}_{AN}= \frac{\mathbf{B}((\mathbf{B}^H\mathbf{B}+\gamma_{AN} \mathbf{I}_N )^{-1})^{\ast}\mathbf{B}^{H}\mathbf{h}(\theta_E, R_E)}{\mathbf{h}^H(\theta_E,R_E)\mathbf{B}((\mathbf{B}^H\mathbf{B}+\gamma_{AN}\mathbf{I}_N)^{-1})^{\ast}\mathbf{B}^H\mathbf{h}(\theta_E,R_E)},
\end{align}
In what follows, in order to make a fair comparison with other methods, $\mathbf{v}_{CM}$ and $\mathbf{v}_{AN}$ can be normalized and denoted as $\mathbf{v}_{CM}(\gamma_{CM})$ and $\mathbf{v}_{AN}(\gamma_{AN})$, respectively. $P_s$ is defined as the total transmit power of Alice and $\beta$ is the parameter that determines the power allocation between the CM and AN. In accordance with the definition of SR, $f_{SR}(\gamma_{CM}, \gamma_{AN} )=\max\{R_d-R_e,~0\}$ can be written as
\begin{align}
&\rm f_{SR}(\gamma_{CM}, \gamma_{AN} ) = \log_2(1+SINR_B)-\log_2(1+SINR_E)\nonumber\\
&= \log_2(1+\frac{g_b\beta P_s|\mathbf{h}^H(\theta_B,R_B)\mathbf{v}_{CM}(\gamma_{CM})|^2}{g_eP_s(1-\beta)|\mathbf{h}^{H}(\theta_B,R_B)\mathbf{v}_{AN}(\gamma_{AN})|^2+\sigma_B^2})\nonumber\\
&-\log_2(1+\frac{g_b\beta P_s|\mathbf{h}^H(\theta_E,R_E)\mathbf{v}_{CM}(\gamma_{CM})|^2}{g_eP_s(1-\beta)|\mathbf{h}^{H}(\theta_E,R_E)\mathbf{v}_{AN}(\gamma_{AN})|^2+\sigma_E^2}).
\end{align}
Obviously, SR is a 2D function of variables $\gamma_{CM}$ and $\gamma_{AN}$. The optimal values of $\gamma_{CM}$ and $\gamma_{AN}$ will be obtained by 2D exhaustive search in the next section.

\section{Simulations}
In our simulation, system parameters are set as follows: quadrature phase shift keying (QPSK) modulation, the total signal bandwidth is 5MHz, $f_c$ = 3GHz, the number of total subcarriers $N_S = 1024$, $d= \lambda /2$, $\beta =0.5$, $\sigma_B^2=\sigma_E^2$ = -60dBm, $(\theta_B, R_B) = (70^{\circ}, 1000 \rm{m})$, and $(\theta_B, R_B) = (100^{\circ}, 750 \rm{m})$.

\begin{figure}[h]
  \centering
  \includegraphics[width=0.4\textwidth]{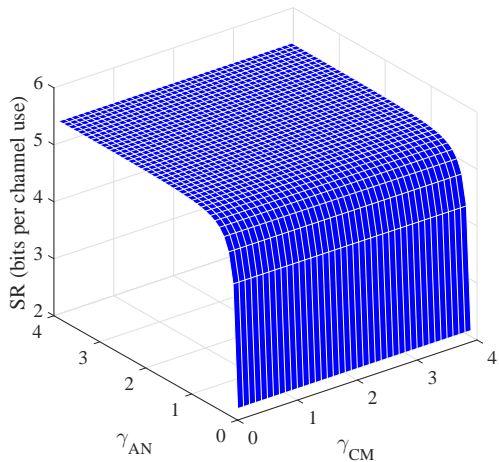}\\
  \caption{3-D  surface of SR versus $\gamma_{CM}$ and $\gamma_{AN}$ with SNR = 20dB and N = 8.
  }\label{gamma}
\end{figure}

 Fig.~\ref{gamma} demonstrates the 3D surface of $f_{SR}(\gamma_{CM}, \gamma_{AN}$ versus $\gamma_{CM}$ and $\gamma_{AN}$.  From Fig.~\ref{gamma}, it can be seen that the SR will reach a flat SR ceil, i.e., the maximum value after  $\gamma_{CM} \ge  0.1$ and $\gamma_{AN} \ge 1.4$. In the following simulation, $\gamma_{CM}$ and $\gamma_{AN}$ are taken to be 2.1 and  1.8.

\begin{figure*}[hpt]
\centering
\subfigure[EA.]{
\begin{minipage}[t]{0.33\linewidth}
\centering
\includegraphics[width=6cm]{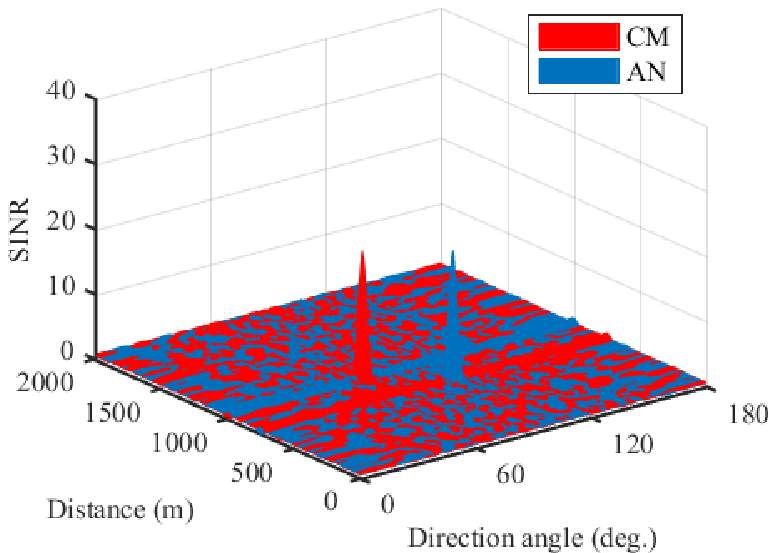}
\end{minipage}%
}%
\subfigure[Proposed Min-TP.]{
\begin{minipage}[t]{0.33\linewidth}
\centering
\includegraphics[width=6cm]{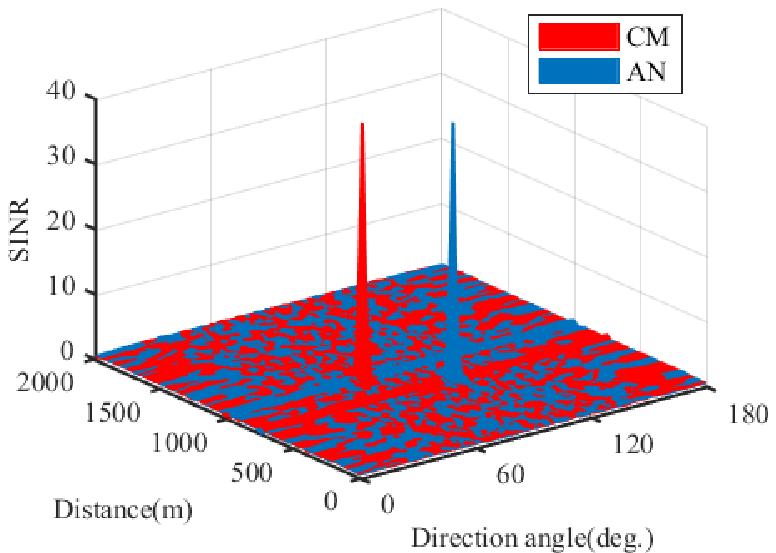}
\end{minipage}%
}%
\subfigure[Proposed Min-RTP.]{
\begin{minipage}[t]{0.33\linewidth}
\centering
\includegraphics[width=6cm]{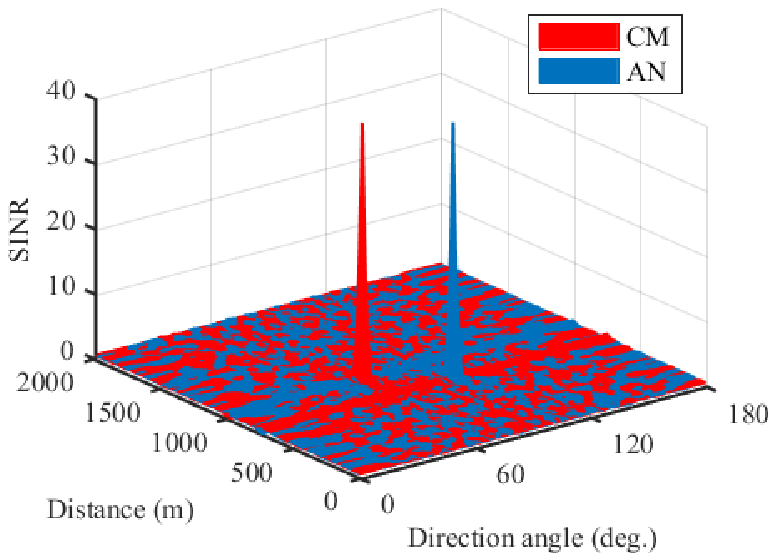}
\end{minipage}%
}%
\centering
\caption{ 3-D  surface of SINR versus direction angle and distance with N = 32.
}\label{SINR}
\end{figure*}

Fig.~\ref{SINR} plots the 3D performance surface of signal-to-interference-plus-noise ratio (SINR) versus direction angle $\theta$ and distance $R$ of the proposed methods for SNR=20dB with EA as a performance reference. Observing three subfigures in Fig.~\ref{SINR}, all methods form  their CM main peaks at Bob  and their AN main peaks at Eve. This is mainly due to the PAC. Clearly, the main peaks of the proposed methods are much higher than those of EA. This result means that they achieve a better SINR performance compared with EA.

\begin{figure}[h]
  \centering
  \includegraphics[width=0.4\textwidth]{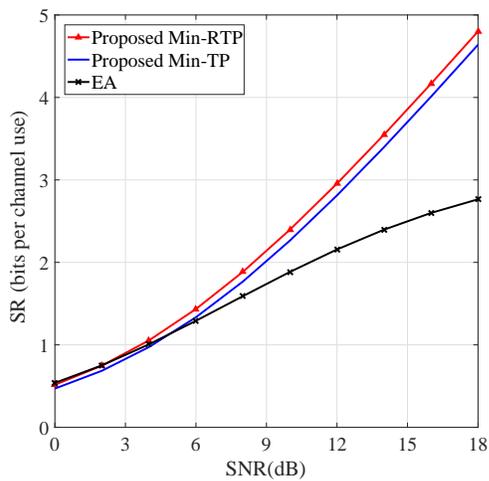}\\
  \caption{ SR versus SNR with N = 8.
}\label{SR}
\end{figure}
Fig.~\ref{SR} illustrates the curves of SR versus SNR of the proposed methods with EA as a performance benchmark. From Fig.~\ref{SR}, it follows that the proposed Min-TP and Min-RTP are  better than EA  in terms of SR in the medium and high SRN regions. As SNR increases, the SR gains achieved by them over EA become more significant. However, in the low SNR region, the three methods have almost the same SR performance.

\begin{figure}[h]
  \centering
  \includegraphics[width=0.4\textwidth]{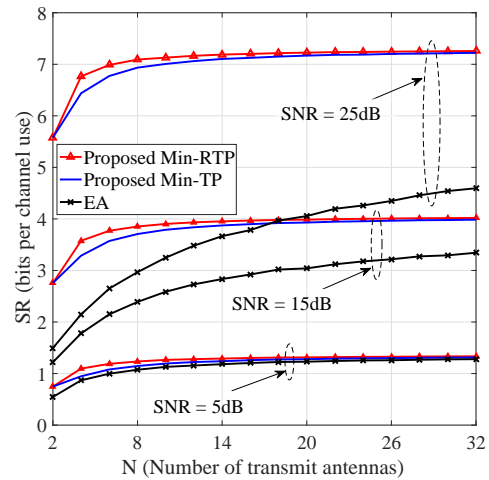}\\
  \caption{ SR versus number of transmit antennas N with three different SNR.
}\label{SR-N}
\end{figure}
Fig.~\ref{SR-N} shows the SR versus the number $N$ of transmit antennas with three different SNR (i.e., 5dB, 10dB, and 20dB). For SNR=5dB, the SR performance difference among the three methods is trivial. For SNR=15dB and SNR=25dB, the SR performance  gains achieved by the proposed methods  become larger.  This outcome is consistent with that in Fig.~\ref{SR}. More importantly, the proposed Min-RTP performs better than Min-TP as the number of antennas at Alice ranges from 2 to 20. Outside the interval, they show an identical SR performance for all three SNRs.

\begin{figure}[h]
  \centering
  \includegraphics[width=0.4\textwidth]{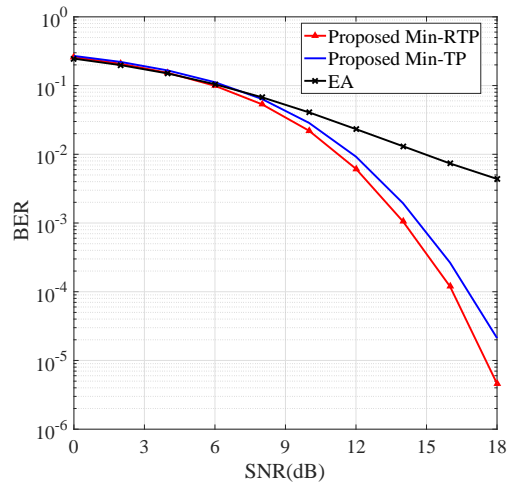}\\
  \caption{ BER versus SNR with N = 8.
  }\label{BER}
\end{figure}
Fig.~\ref{BER} shows the curves of BER versus  SNR for the  proposed Min-RTP and Min-TP with EA as a performance reference. According to this figure, it is evident that the BER performance of our proposed schemes outperforms that of EA for SNR$\ge 6dB$.  Their BER performance has an increasing order as follows: EA, Min-TP, and Min-RTP.

\section{Conclusion}
In this paper, two efficient beamforming methods: Min-TP and Min-RTP, have been proposed to achieve SPJC. The latter is proposed by introducing regularized penalty to address the singular problem of the former. Simulation results show that the proposed Min-TP and Min-RTP scheme perform better than EA in terms of BER and SR in the medium and high SNR regions. The proposed Min-RTP is slightly better than the proposed Min-TP in terms of SR and BER.

\ifCLASSOPTIONcaptionsoff
  \newpage
\fi
\bibliographystyle{IEEEtran}
\bibliography{IEEEfull,cite}

\begin{thebibliography}{10}
\providecommand{\url}[1]{#1}
\csname url@samestyle\endcsname
\providecommand{\newblock}{\relax}
\providecommand{\bibinfo}[2]{#2}
\providecommand{\BIBentrySTDinterwordspacing}{\spaceskip=0pt\relax}
\providecommand{\BIBentryALTinterwordstretchfactor}{4}
\providecommand{\BIBentryALTinterwordspacing}{\spaceskip=\fontdimen2\font plus
\BIBentryALTinterwordstretchfactor\fontdimen3\font minus
  \fontdimen4\font\relax}
\providecommand{\BIBforeignlanguage}[2]{{%
\expandafter\ifx\csname l@#1\endcsname\relax
\typeout{** WARNING: IEEEtran.bst: No hyphenation pattern has been}%
\typeout{** loaded for the language `#1'. Using the pattern for}%
\typeout{** the default language instead.}%
\else
\language=\csname l@#1\endcsname
\fi
#2}}
\providecommand{\BIBdecl}{\relax}
\BIBdecl

\bibitem{wyner}
A.~D. {Wyner}, ``The wire-tap channel,'' \emph{Bell. Syst. Tech. J.}, vol.~54,
  no.~8, pp. 1355--1387, Oct. 1975.

\bibitem{wanghuiming}
H.~{Wang}, Q.~{Yin}, and X.~{Xia}, ``Distributed beamforming for physical-layer
  security of two-way relay networks,'' \emph{IEEE Trans. Signal Process.},
  vol.~60, no.~7, pp. 3532--3545, Jul. 2012.

\bibitem{zhaonan}
N.~{Zhao}, F.~R. {Yu}, M.~{Li}, and V.~C.~M. {Leung}, ``Anti-eavesdropping
  schemes for interference alignment (ia)-based wireless networks,'' \emph{IEEE
  Trans. Wireless Commun.}, vol.~15, no.~8, pp. 5719--5732, Aug. 2016.

\bibitem{chenxiaoming}
X.~{Chen}, D.~W.~K. {Ng}, W.~H. {Gerstacker}, and H.~{Chen}, ``A survey on
  multiple-antenna techniques for physical layer security,'' \emph{IEEE Commun.
  Surv. Tut.}, vol.~19, no.~2, pp. 1027--1053, Secondquarter 2017.

\bibitem{wuyongpeng}
Y.~{Wu}, J.~{Wang}, J.~{Wang}, R.~{Schober}, and C.~{Xiao}, ``Secure
  transmission with large numbers of antennas and finite alphabet inputs,''
  \emph{IEEE Trans. Commun.}, vol.~65, no.~8, pp. 3614--3628, Aug. 2017.

\bibitem{wangzhengwang}
F.~{Shu}, Z.~{Wang}, R.~{Chen}, Y.~{Wu}, and J.~{Wang}, ``Two high-performance
  schemes of transmit antenna selection for secure spatial modulation,''
  \emph{IEEE Trans. Veh. Technol.}, vol.~67, no.~9, pp. 8969--8973, Sep. 2018.

\bibitem{xiaguiyang}
G.~{Xia}, F.~{Shu}, Y.~{Zhang}, J.~{Wang}, S.~{ten Brink}, and J.~{Speidel},
  ``Antenna selection method of maximizing secrecy rate for green secure
  spatial modulation,'' \emph{IEEE Trans. Green Commun. Netw.}, vol.~3, no.~2,
  pp. 288--301, Jun. 2019.

\bibitem{yuxiangbin}
X.~{Yu}, Y.~{Hu}, Q.~{Pan}, X.~{Dang}, N.~{Li}, and M.~H. {Shan}, ``Secrecy
  performance analysis of artificial-noise-aided spatial modulation in the
  presence of imperfect csi,'' \emph{IEEE Access}, vol.~6, pp.
  41\,060--41\,067, 2018.

\bibitem{yuanding}
Y.~{Ding} and V.~F. {Fusco}, ``A vector approach for the analysis and synthesis
  of directional modulation transmitters,'' \emph{IEEE Trans. Antennas
  Propag.}, vol.~62, no.~1, pp. 361--370, Jan. 2014.

\bibitem{Zhou2019UAV}
X.~{Zhou}, Q.~{Wu}, S.~{Yan}, F.~{Shu}, and J.~{Li}, ``{UAV}-enabled secure
  communications: Joint trajectory and transmit power optimization,''
  \emph{IEEE Trans. Veh. Technol.}, vol.~68, no.~4, pp. 4069--4073, Apr. 2019.

\bibitem{huAN}
J.~{Hu}, S.~{Yan}, F.~{Shu}, J.~{Wang}, J.~{Li}, and Y.~{Zhang},
  ``Artificial-noise-aided secure transmission with directional modulation
  based on random frequency diverse arrays,'' \emph{IEEE Access}, vol.~5, pp.
  1658--1667, 2017.

\bibitem{wuxiaomin1}
F.~{Shu}, X.~{Wu}, J.~{Hu}, J.~{Li}, R.~{Chen}, and J.~{Wang}, ``Secure and
  precise wireless transmission for random-subcarrier-selection-based
  directional modulation transmit antenna array,'' \emph{IEEE J. Sel. Areas
  Commun.}, vol.~36, no.~4, pp. 890--904, Apr. 2018.

\bibitem{linjingran}
J.~{Lin}, Q.~{Li}, J.~{Yang}, H.~{Shao}, and W.~{Wang}, ``Physical-layer
  security for proximal legitimate user and eavesdropper: A frequency diverse
  array beamforming approach,'' \emph{IEEE Trans. Inf. Forensics Security},
  vol.~13, no.~3, pp. 671--684, Mar. 2018.

\bibitem{qiubin}
B.~{Qiu}, J.~{Xie}, L.~{Wang}, and Y.~{Wang}, ``Artificial-noise-aided secure
  transmission for proximal legitimate user and eavesdropper based on frequency
  diverse arrays,'' \emph{IEEE Access}, vol.~6, pp. 52\,531--52\,543, 2018.

\bibitem{shentong:journals/corr/abs-1808-01896}
\BIBentryALTinterwordspacing
T.~Shen, S.~Zhang, R.~Chen, J.~Wang, J.~Hu, F.~Shu, and J.~Wang, ``Two
  practical random-subcarrier-selection methods for secure precise wireless
  transmission,'' \emph{CoRR}, vol. abs/1808.01896, 2018. [Online]. Available:
  \url{http://arxiv.org/abs/1808.01896}
\BIBentrySTDinterwordspacing

\bibitem{Qinyaolu}
F.~{Shu}, Y.~{Qin}, T.~{Liu}, L.~{Gui}, Y.~{Zhang}, J.~{Li}, and Z.~{Han},
  ``Low-complexity and high-resolution doa estimation for hybrid analog and
  digital massive mimo receive array,'' \emph{IEEE Trans. Commun.}, vol.~66,
  no.~6, pp. 2487--2501, Jun. 2018.

\end{thebibliography}
\end{document}